# Blue to green single photons from InGaN/GaN dot-in-a-nanowire ordered arrays

E. Chernysheva[1], Z. Gačević[2], N. García-Lepetit[2], H. P. van der Meulen[1], M. Müller[3], F. Bertram[3], P. Veit[3], A. Torres-Pardo[4], J. M. González Calbet[4], J. Christen[3], E. Calleja[2], J.M. Calleja[1] and S. Lazić[1(a)]

[1] *Departamento de Física de Materiales, Instituto Nicolás Cabrera and Instituto de Física de Materia Condensada (IFIMAC), Universidad Autónoma de Madrid - Francisco Tomás y Valiente 7, 28049 Madrid, Spain*
[2] *ISOM-DIE, Universidad Politécnica de Madrid - Avenida Complutense 30, 28040, Madrid, Spain*
[3] *Institute of Experimental Physics, Otto-von-Guericke-University - Universitätsplatz 2, 39106 Magdeburg, Germany*
[4] *Centro Nacional de Microscopía Electrónica, Universidad Complutense de Madrid - Avenida Complutense s/n, 28040, Madrid, Spain*

**Abstract** – Single photon emitters (SPEs) are at the basis of many applications for quantum information management. Semiconductor-based SPEs are best suited for practical implementations because of high design flexibility, scalability and integration potential in practical devices. Single photon emission from ordered arrays of InGaN nano-disks embedded in GaN nanowires is reported. Intense and narrow optical emission lines from quantum dot-like recombination centers are observed in the blue-green spectral range. Characterization by electron microscopy, cathodoluminescence and micro-photoluminescence indicate that single photons are emitted from regions of high In concentration in the nano-disks due to alloy composition fluctuations. Single photon emission is determined by photon correlation measurements showing deep antibunching minima in the second order correlation function. The present results are a promising step towards the realization of on-site/on-demand single photon sources in the blue-green spectral range operating in the GHz frequency range at high temperatures.

**Introduction.** – Single photons are ideal "flying" qubits to convey quantum information between distant nodes of a quantum network. Reliable and controlled generation of single photons is therefore a crucial step to develop applications for quantum communication, quantum information processing and quantum metrology [1,2]. Single photons can be emitted in principle by material entities possessing discrete energy levels, as they need a finite time to "recharge" after emission of one photon. The standard method to assess single photon emission is to measure the second order photon correlation function by Hanbury-Brown and Twiss (HBT) interferometry. As shown in Fig. 1, single photons are either reflected or transmitted by a beam splitter, so that the probability of simultaneous detection in the two detectors of the interferometer is zero. The detection events are stored in a Time-Correlated Single Photon Counter (TCSPC), and the resulting correlation function $g^2(\tau)$ shows an "antibunching" dip at zero delay time between the detectors.

Various quantum systems have been actively explored for the realization of SPEs, including atoms, ions defect centers in solids and quantum dots [3-6 and references therein]. In particular, semiconductor quantum dots (QDs) [7,8] have emerged as promising candidates for practical SPEs due to their long-term mechanical, chemical and optical stability, small emission linewidth, fast radiative lifetime and the possibility of electrical operation [9-11]. Besides, their fabrication is compatible with present semiconductor technologies allowing their integration into existing photonic and optoelectronic circuits. Wide bandgap semiconductor QDs based on group III-nitrides present additional advantages, as high-temperature operation [11-14] and wide spectral tunability from the deep ultraviolet to telecommunication wavelengths. The III-nitride-based quantum emitters operating in the ultraviolet-visible region are especially beneficial for free space communications [15]. Moreover, highly polarized light emission originating from valence-band mixing effects is typical of III-nitrides. This property is necessary for the implementation of quantum key distribution systems [16] and linear optical quantum computing [17].

E. Chernysheva *et al.*

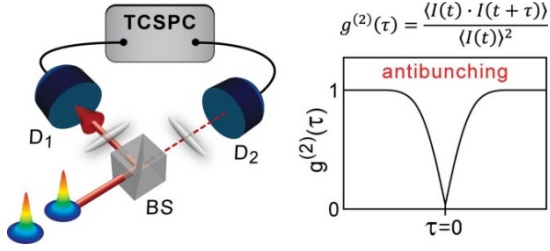

Fig. 1 (Color on-line) Left: HBT setup. Single photons are either reflected or transmitted at the beam splitter (BS) and detected at detector $D_1$ or $D_2$. The TCSPC produces a histogram of detection coincidences as a function of the delay time τ between the detectors. Right: Correlation function as a function of delay time showing the anti-bunching dip.

In spite of significant advances in the field, the development of solid-state systems for quantum information management requires improvements in the control of the SPE location and emission wavelength. The growth of III-nitride self-assembled QDs on planar substrates results in random SPE location. Recently, the inclusion of III-nitride QDs into semiconductor quantum wires has been demonstrated to be an effective approach for controlling the spatial location of single quantum emitters. Several types of nanostructures have been reported [11,18,19], which include ordered arrays of SPEs based on GaN [14] and InGaN [20]. Among these, the III-nitride nanowire heterostructures are especially appealing due to their small nanowire-substrate interface, which enables strain-free heteroepitaxy on substrates with large lattice mismatch. The strain-free growth reduces the density of threading dislocations, resulting in high structural and optical quality [21]. Besides, the dot-in-a-nanowire approach helps to improve the emission efficiency and directionality [22,23] allowing the development of highly efficient SPEs [24].

In this work we report on InGaN/GaN single photon emitters in the blue-green spectral range grown by molecular beam epitaxy (MBE) [25]. Compared to the widely used metal-organic vapor-phase deposition technique [14,18-20], MBE allows to increase the In concentration range and thus the emission spectral range of the SPEs. Our samples are formed by InGaN nano-disks embedded in GaN nanowires arranged in periodic two-dimensional arrays. The single photon emission is demonstrated by strong antibunching in photon correlation measurements and displays almost full linear polarization. It originates from QD-like centers in the nanowire InGaN sections, as demonstrated by scanning transmission electron microscopy (STEM) and cathodo-luminescence (CL) measurements. The formation of such QD-like centers is evidenced by sub-meV emission linewidths in micro-photoluminescence (μ-PL) spectra. Single photon emission in the green region based on nitride systems was so far only reported in MBE-grown self-assembled nanowire-QDs by S. Deshpande et al. [11]. The present results are, therefore, an important step towards the realization of on-site/on-demand single photon sources operating up to GHz frequencies in a broad range of the visible spectrum.

**Experiment.** – Periodic GaN nanowire arrays hosting InGaN nano-disks were grown by plasma-assisted molecular beam epitaxy (PA-MBE) on (0001) GaN templates [26,27]. A titanium nano-hole mask covering large sample areas was fabricated in a compact hexagonal lattice using colloidal lithography [28]. Periodic arrays of InGaN/GaN disk-in-nanowire heterostructures with 280 nm pitch were then obtained by selective area growth (SAG). Typical nanowire diameter and height values are about 200 and 500 nm, respectively. Individual nanowires were mechanically removed from the native sample and dispersed on a silicon wafer covered with a titanium metal grid. Both CL and μ-PL techniques were used to study the optical emission of the nanowires. The CL detection unit was integrated in a FEI (S) TEM Tecnai F20 equipped with a liquid helium stage (T = 10-300 K). The light was collected by a retractable parabolic aluminum mirror, dispersed in a grating monochromator (MonoCL4, Gatan) and detected by a liquid nitrogen-cooled silicon charge coupled device (CCD) camera. The material contrast at each position was recorded simultaneously to the detection of the CL-signal: Electrons that are forward scattered into high as well as small solid angles are acquired by an annular dark-field detector (HA)ADF from Fischione (model 3000). The STEM acceleration voltage was optimized to 80 kV to minimize sample damage. The μ-PL experiments were performed on samples mounted in a continuous flow helium cryostat at 10 K. A continuous-wave helium-cadmium (He-Cd) laser operating at 325 and 442 nm was used as excitation source. The laser beam was focused onto a ~1.5 μm spot on the sample through a 100x microscope objective (NA=0.73). The emitted photons were collected by the same objective and detected by a single grating monochromator with a spectral resolution of ~ 350 μeV and a nitrogen-cooled CCD camera.

For photon correlation measurements, a HBT interferometer [29] was placed on the side exit of the monochromator. The detection efficiency of the detectors (avalanche photodiodes or APDs) in the measured spectral range was ~35% and their response time was $\tau_{IRF}$ ~ 350 ps. A short pass filter in front of one of the APDs suppressed the optical cross-talk between the two APDs. The signal from the detectors was sent to a time-correlated single photon counting system, where correlation histograms as a function of the time delay between the detectors were obtained. The histograms were acquired with 32 ps time bin and count rates up to $1.5\times10^5$ photons/s. All HBT measurements were performed at 10 K.

**Results and discussion.** – The scanning electron microscopy (SEM) image of a typical nanowire array with areal density of ~$1.6\times10^9$ cm$^{-2}$ is shown in Fig. 2(a). The nanowires exhibit hexagonal cross section with lateral facets defined by non-polar m-planes. The specific growth conditions (temperature and ingredient fluxes) determine the shape of the InGaN nano-disks embedded inside the nanowire tip [27]. As shown in the STEM micrograph of an individual nanowire (Fig. 2(b)), a ~30 nm thick InGaN region with trapezoidal lateral cross section is formed on the nanowire top facet, together with narrower InGaN regions at the lateral semi-polar facets. The exact width of the InGaN side region is difficult to determine from the STEM images as they are taken across the entire nanowire (i.e. internal InGaN side-facets are less visible than the external ones). The CL images in Fig. 2(c) show the spatial and spectral distribution of the light emitted from a single nanowire, recorded at three different heights, as marked by the horizontal lines in the right panel of Fig. 2(b). The left panel in Fig. 2(c) corresponding to a region below the InGaN nanodisks (line 1) is dominated by the bulk GaN exciton at 3.45 eV. In the central panel, emission from the lateral parts of the InGaN nanodisk (line 2) is observed at 3.06 eV, while the right panel shows the emission from the top central part of the nanodisks (line 3) at 2.45 eV. The energy difference between the two InGaN emissions is attributed to a higher In incorporation rate at the polar c-plane [30], as well as to the stronger electric field [31], as compared to the semi-polar r-planes. Note also the spectral width of the InGaN emissions (~100 meV), which is due to random In content fluctuations inside a single nanodisk.

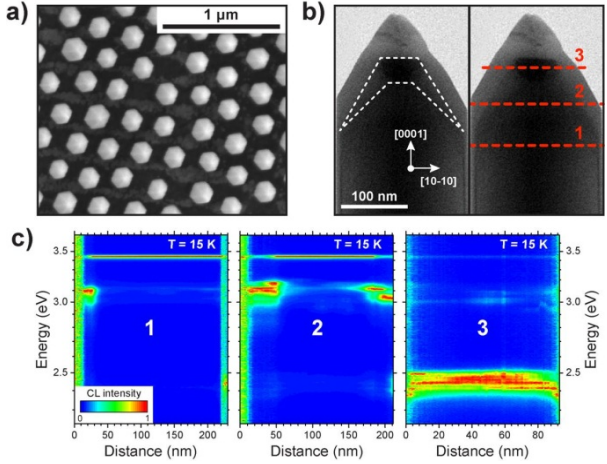

Fig. 2 (Colour on-line) a) Top view SEM image of a sample showing an array of InGaN/GaN disk-in-nanowire heterostructures. b) Cross-sectional STEM bright-field image of a single nanowire showing the embedded InGaN nanodisk. c) CL line scans recorded at heights 1 to 3, as shown in b).

The same emission bands are observed in the μ-PL spectra of the nanowires shown in Fig. 3 (a) for an "as grown" sample. No yellow-band emission is observed, contrary to what is typically found in other GaN-based systems. The light originating from the central part of the InGaN disks (Fig. 3(b)) can be tuned across the blue-green spectral emission region (~2.80 to 2.35 eV) by varying the growth temperature and In flux. The emission of a single dispersed nanowire is presented in Fig. 3(c). The spectral range of the lowest energy emission of both "as-grown" and dispersed nanowires (~100 meV) is similar in both cases to the one observed in CL (Fig. 2(c)), indicating that the In content fluctuations are similar in different nanowire samples. The intense, sharp lines observed in the μ-PL spectra of both "as-grown" and dispersed nanowires indicate that some of these fluctuations are strong enough to produce quantum confinement of excitons. These QD-like centers are, in fact, the single photon emitters discussed later. The linewidth of the QD-like emission (350-600 μeV, as shown in the inset of Fig. 3(c)), is similar to values reported for high quality self-assembled [13,32] and SAG [19] InGaN QDs.

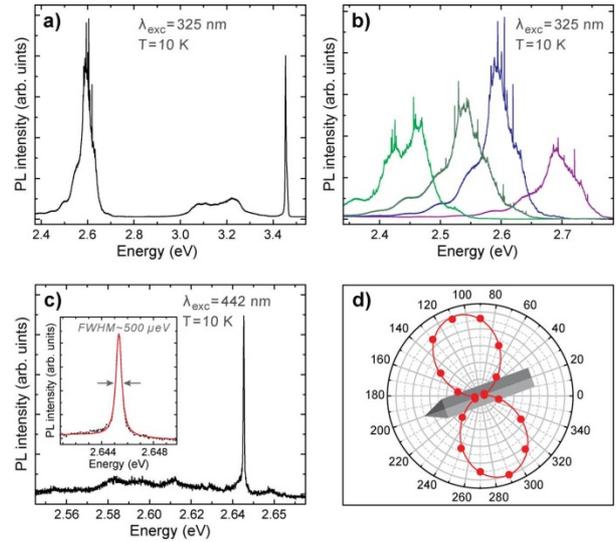

Fig. 3 (Color on-line) a) μ-PL spectrum of an "as grown" sample. b) Enlarged spectra of the QD-like emission region for different In concentrations in the InGaN nanodisks. c) μ-PL spectrum of a single dispersed nanowire from the same batch as in a). The inset in c) shows an enlarged QD-like line with a Lorentzian fit (red line). d) Angular polarization distribution of the QD-like emission, showing strong linear polarization perpendicular to the c-axis.

The QD transitions are visible at temperatures up to ~80 K (not-shown). No indications of spectral diffusion or blinking were observed in the present case. The linear dependence of the integrated PL peak intensity on the excitation power density (not shown), as well as its Lorentzian line-shape (inset in Fig. 3(c)), confirm that the sharp PL lines are due to recombination of excitons confined in QD-like potential minima due to the formation of In-rich clusters. In this context the emission background observed between 2.55 and 2.65 meV in Fig. 3(c) for a single dispersed nanowire is attributed to InGaN regions without QD-like confinement. A weak biexciton emission is observed in some of the nanowires at energy higher than the exciton emission. The negative biexciton binding energy (~ -20 meV) is indicative of strong confinement in our QD-like centers. All the QD-like emission lines present a high linear polarization degree (> 90%), as shown in Fig. 3(d). This is crucial for quantum information applications where polarization controlled single photon emission is required

E. Chernysheva *et al.*

[16,17]. The linear polarization is ubiquitous in InGaN-based QDs and results from the valence band mixing induced by the in-plane anisotropy of strain, as well as QD shape [13,19,33,34]. It should be noted that no preferred crystallographic direction of the polarization was found, thus suggesting random asymmetry of the nanowire QD structures.

Single photon emission from the InGaN nano-disk is demonstrated by photon correlation measurements. The normalized number of coincidence counts for the single dispersed nanowire shown in Fig. 3(c) is presented in Fig. 4 as a function of the delay time between the two APDs of the HBT interferometer. The red curve is a fit of the data with the

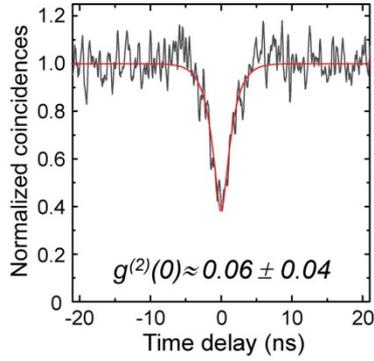

Fig. 4 (Colour on-line) Normalized coincidence counts for the QD emission peak in Fig. 3 c) showing the antibunching dip. The red line is a fit with Eq. (1), convolved with Eq. (2).

uncorrected second order correlation function:

$$g_e^{(2)}(\tau) = 1 - \beta \cdot e^{-|\tau/\tau_R|} \quad (1)$$

where $\tau_R$ is the anti-bunching time and $\tau$ is the time delay, convoluted with the instrumental response function [10]:

$$h(\tau) = C \cdot e^{-|\tau/\tau_{IRF}|} \quad (2)$$

where C is a constant and $\tau_{IRF} \sim 350$ ps. For an ideal single photon emitter [35], the coincidence counts must vanish at zero delay time, corresponding to $\beta = 1$. The coincidence counts in Fig. 4 show a pronounced anti-bunching dip at $\tau = 0$ ($\beta = \sim 0.6$). After correction by the instrumental response time one obtains: $g^{(2)}(0) = 0.2 \pm 0.04$. Besides, the data must also be corrected for the background counts [10], which include the detectors dark counts and the contribution from additional emission centers within the detection area:

$$g^{(2)}(0) = [g_e^{(2)}(0) - (1 - \rho^2)]/\rho^2 \quad (3)$$

where $\rho=S/(S+B)$ is the signal-to-total counts ratio and S (B) stands for the signal (background) counts. In the case of Fig. 4, a value of $\rho = 0.92$ is used, estimated from the emission spectrum recorded using the HBT-APDs. After background correction we obtain $g^{(2)}(0) = 0.06 \pm 0.04$, indicating that the QDs present in our InGaN nano-disks would be close to ideal SPEs, if the background could be suppressed. Nanowires in the "as-grown" sample also present strong anti-bunching, although background corrections in this case are more important. The anti-bunching time can also be extracted from the fit with the convolution of equations (1) and (2).

Changes in the emission dynamics under different optical excitation conditions affect the anti-bunching rate $1/\tau_R$, according to [34]:

$$\frac{1}{\tau_R} = \frac{1}{\tau_X} + \gamma \quad (4)$$

where $\tau_X$ is the exciton decay time and $\gamma$ is the pump rate. The coincidence counts histogram was recorded for different values of the excitation power, as shown in Fig. 5(a), where an apparent decrease of the anti-bunching time with increasing power is observed. The anti-bunching rates deduced from these measurements are plotted in Fig. 5(b) as a function of the optical pump power. It follows the expected linear trend of Eq. (4), extrapolating to $1/\tau_X = 0.84 \pm 0.10$ ns$^{-1}$ at zero excitation power. This value coincides with the exciton decay rate measured independently by time-resolved PL: $1/\tau_X = 0.74 \pm 0.02$ ns$^{-1}$. Such exciton decay rate is comparable with that typical for polar InGaN QDs emitting in the same range [11] and is suitable for the realization of triggered single photon sources driven at frequencies up to a GHz range. An estimate of the absorption cross section $\sigma_a$ of our SPEs can be obtained from the anti-bunching time dependence on the excitation intensity ($P_{exc}$). From the slope of the red line in Fig. 3(b) and the expression of the pump rate in terms of the excitation power and the photon energy hv: $\gamma = \sigma_a P_{exc}/h\nu$ we obtain an absorption cross section of $\sigma_a = 1.7 \times 10^{-14}$ cm$^2$. This value is comparable to that of colloidal CdSe QDs [36] and one to two-orders of magnitude larger than in epitaxial InAs QDs [37].

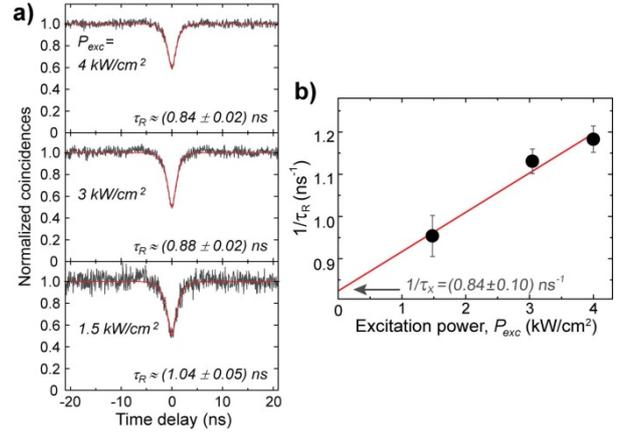

Fig. 5 (Colour on-line) a) Evolution of the normalized coincidence count distribution with the excitation power. The corresponding anti-bunching times are extracted from the fits (see text). b) Dependence of the anti-bunching rate on the optical excitation power (full dots) and linear fit (red line) to Eq. (4). The extrapolation to vanishing excitation power corresponds to the exciton decay rate $1/\tau_X$.

**Conclusion.** – In summary, we have demonstrated the feasibility of MBE-grown ordered arrays of quantum emitters based on InGaN nano-disks embedded in GaN nanowires. Epitaxial growth by selective area PA-MBE permits to fabricate the InGaN/GaN disk-in-nanowire heterostructures at predetermined positions as well as to tune their emission across the blue-green spectral region. Non-classical light is emitted from QD-like centers located in the apices of the InGaN nanodisks, which show sharp and linearly polarized emission lines. The anti-bunching dips in photon correlation measurements are clear signature of single photon operation, while the linear dependence of the anti-bunching rate on excitation power allows to determine the maximum single photon emission rate. Our results represent a promising step towards the realization of on-demand SPEs operating at high repetition rates, which is important for future high-speed quantum information processing. These nanowire heterostructures could also serve as fundamental building blocks for SPE integration into on-chip quantum photonic circuits, since they can be easily transferred from its native to a foreign substrate.

***

This work was supported by the Spanish MINECO under contracts MAT2011-22997and MAT2014-53119-C2-1-R. E.C. acknowledges the FPI grant (MINECO). The authors thank David Lopez Romero for the help in metal mask fabrication and Dr. Ana Bengoechea-Encabo and Steven Albert for the help in lithographic mask preparation and MBE growth. M.M., F.B., P.V. and J.C. thank the German DFG for financial support (research program INST 272/148-1) and the collaborative research center SFB 787 "Semiconductor Nanophotonics: Materials, Models and Devices".